\documentclass{article}
\usepackage{arxiv}

\usepackage[utf8]{inputenc} % allow utf-8 input
\usepackage[T1]{fontenc}    % use 8-bit T1 fonts
\usepackage[bookmarks=false]{hyperref}
\usepackage[table]{xcolor}
\usepackage[toc,page]{appendix}
\usepackage{pdfpages}
\usepackage{float}
\usepackage{booktabs}
\usepackage{adjustbox}
\usepackage{caption} 
\usepackage{subcaption}
\usepackage{mathtools}
\usepackage{biblatex}
\title{Deep Learning modelling of the Limit Order Book: \\ a comparative perspective}

\author{
  Antonio Briola\\
  Department of Computer Science\\
  UCL, London, United Kingdom\\
  \texttt{a.briola@ucl.ac.uk} \\
  \and
  \textbf{Jeremy Turiel}\\
  Department of Computer Science\\
  UCL, London, United Kingdom\\
  \texttt{jeremy.turiel.18@ucl.ac.uk} \\
  \And
  \textbf{Tomaso Aste}\\
  Department of Computer Science\\
  UCL, London, United Kingdom\\
  and \\
  Systemic Risk Centre\\
  London School of Economics, London, United Kingdom\\
  \texttt{t.aste@ucl.ac.uk} \\
}

\begin{document}
\maketitle

\begin{abstract}

    The present work addresses theoretical and practical questions in the domain of Deep Learning for High Frequency Trading. 
    State-of-the-art models such as Random models, Logistic Regressions, LSTMs, LSTMs equipped with an Attention mask, CNN-LSTMs and MLPs are reviewed and compared on the same tasks, feature space, and dataset and clustered according to pairwise similarity and performance metrics. The underlying dimensions of the modelling techniques are hence investigated to understand whether these are intrinsic to the Limit Order Book's dynamics.
    We observe that the Multilayer Perceptron performs comparably to or better than state-of-the-art CNN-LSTM architectures indicating that dynamic spatial and temporal dimensions are a good approximation of the LOB's dynamics, but not necessarily the true underlying dimensions.

\end{abstract}


\section{Introduction} \label{intro}

Recent years have seen the growth and spreading of Deep Learning methods across several domains. In particular, Deep Learning has been increasingly applied to the domain of Financial Markets. However, these activities are mostly performed in industry and there is a scarce academic literature to date. The present work builds upon the general Deep Learning literature to offer a comparison between models applied to High Frequency markets. Insights about Market Microstructure are then derived from the features and performance of the models.

\begin{figure}[H]
    \centering
    \includegraphics[scale=0.28]{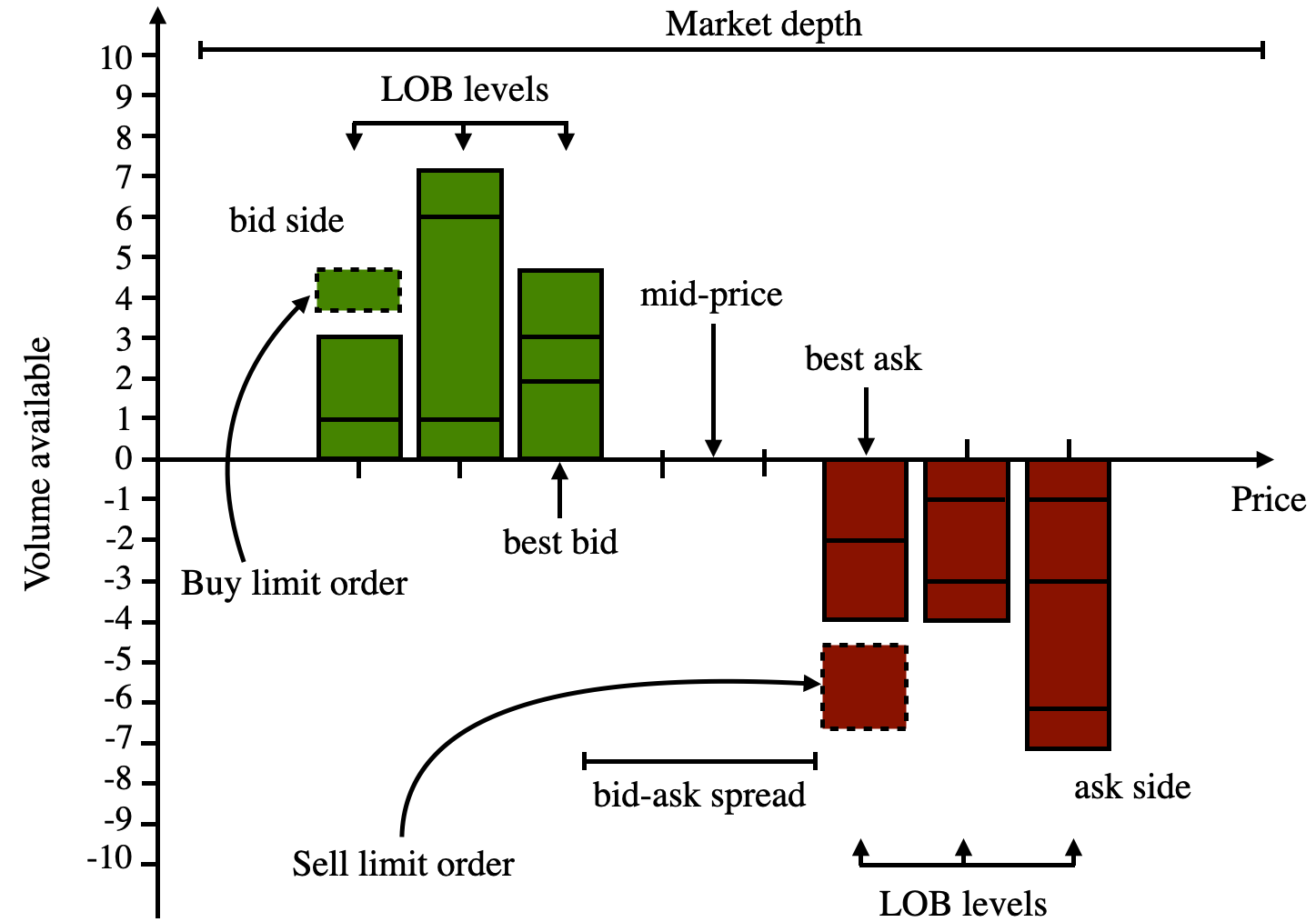}
    \caption{Schematic representation of the LOB structure. It is possible to distinguish between the bid side (left) and the ask side (right), where both are organised into levels. The first level contains the best bid-price and the best ask-price, respectively. Since the market's goal is to facilitate the matching of intentions from buyers and sellers, the best bid-price is defined as the maximum proposed bid price, while the best-ask price is defined as the minimum proposed ask-price. The distance between best bid-price and best ask-price is commonly referred to as bid-ask spread. The mid-price is defined as the mean between best bid-price and best ask-price. The lower (higher) the bid-price (ask-price) at which limit orders are submitted, the deeper the level at which they are placed. The cumulative volume of buy and sell limit orders determines the market depth. In order-driven markets, the priority of orders to be matched at each price level depends upon the arrival time, according to a \textit{FIFO} (First In, First Out) rule \cite{abergel2016limit}.}
    \label{LOB_Schema_Figure}
\end{figure}

The Limit Order Book (LOB) represents the venue where buyers and sellers interact in an order-driven market. It summarises a collection of intentions to buy or sell integer multiples of a base unit volume $v_0$ (lot size) at price $p$. The set of available prices $\{p_0, ..., p_n\}$ is discrete with a basic unit step $\theta$ (tick size). The LOB is a self-organising complex process where a transaction price emerges from the interaction of a multitude of agents in the market. These agents interact through the submission of a range of order types in the market. Figure \ref{LOB_Schema_Figure} provides a visual representation of the LOB, its components and features.

%is discrete with a fixed tick size $\theta$ defining the basic unit step. The LOB is a self-organising complex process where a transaction price emerges from the interaction of a multitude of agents in the market. 

%The main categories of orders are
Three main categories of orders exist: market orders, limit orders and cancellation orders. Market orders are executed at arrival by paying a higher execution cost which originates from crossing the bid-ask spread. Limit orders make up the liquidity of the LOB at different price levels and constitute an expression of the intent to buy or sell a given quantity $v_p$ at a specific price $p$. These entail lower transaction costs, with the risk of the order not being fulfilled. Cancellation orders are used to partially or fully remove limit orders which have not been filled yet.

The study of order arrival and dynamics of the Limit Order Book and of order-driven markets has seen a growing interest in the academic literature as well as in the industry. This sparked from the almost simultaneous spreading of electronic trading and high frequency trading (HFT) activity throughout global markets. The resulting increase in frequency of the trading activity has generated a growing amount of trading data thereby creating the critical mass for Big Data applications.

The availability of Big Data from High Frequency Trading has then made it possible to apply the data hungry Machine Learning and Deep Learning methods to financial markets.
Machine Learning methods were initially adopted by hedge funds towards the end of the last century, while now adoption is expanding and it is possible to see large quantitative firms and leading investment banks openly applying AI methods. Building upon this growing interest, an increasing number of papers and theses exploring Machine Learning and Deep Learning methods applied mostly to financial markets are being written. This is part of the modern trend where large companies lead research fields in AI due to the availability of computational and monetary resources \cite{ganesh2019reinforcement, Li_Saude}. This often results in a literature dominated by increasingly complex and task-specific model designs, often conceived adopting an applicative approach without an in-depth analysis of the theoretical implications of obtained results.

In light of this, for the present work, the relevant literature has been screened in search of state-of-the-art models for price movement forecasting in high frequency trading. Increasingly complex models from this literature are presented, characterised and results are compared on the same training and test sets. Theoretical implications of these results are investigated and they are compared for statistically validated similarity. This analysis has the purpose to reason why certain models should or should not be used for this application as well as verify whether more complex architectures are truly necessary. One example of this consists in the performed study of temporal and spatial dimensions (implied by Recurrent Neural Network (RNN) \cite{lipton2015critical}, Attention \cite{vaswani2017attention} and Convolutional Neural Network (CNN) \cite{lecun1995convolutional} models, respectively) and whether they are unique and optimal representations of the Limit Order Book. The alternative hypothesis consists in a Multilayer Perceptron (MLP) which does not explicitly model any dynamic.

The Deep Learning models described in Section \ref{rel_work}, incorporate assumptions about the structure and relations in the data as well as how it should be interpreted. As reported above, this is the case of CNNs which exploit the relations between neighbouring elements in their grid-like input. Analogous considerations can be made for RNN-like models, which are augmented by edges between consecutive observation embeddings. These types of structures aim to carry information across a series, hence implying sequential relations between inputs. These tailored architectures are used to test hypotheses about the existence and informativeness of corresponding dimensional relations in the Limit Order Book.

The diffusion of flexible and extensible frameworks such as Weka \cite{witten2016data} and Keras \cite{keras_description}, and the success of automated Machine Learning tools such as AutoML \cite{automlchallenges} and H2O AutoML \cite{candel2016deep}, is facilitating the abovementioned industry driven applicative approach. Unfortunately, much less attention is given to methods for model comparison able to asses the real improvement brought by these new models \cite{demvsar2008appropriateness}. In order to make the work presented here more reliable and to promote a more thorough analysis of published works, a statistical comparison between models is provided through the state-of-the-art statistical approaches described in \cite{benavoli2017time}.

It is crucial for scientifically robust results to validate the significance of model performance and the potential performance improvements brought by novel architectures and methodologies. The main classes of methods for significance testing are Frequentist and Bayesian. Null model-based Frequentist methods are not ideal for this application as the detected statistical significance might have no practical impact on performance. The need to answer questions about the likelihood of one model to perform significantly better than another, based on experiments requires the use of posterior probabilities, which can be obtained from the Bayesian methods as, for instance, in \cite{edwards1963bayesian, dickey1973scientific, berger1987testing}.

%The comparisons are made as there is no reason, in principle, to assume that the improvement in performance brought by a novel architecture is statistically significant and robust. Null hypothesis-based significance testing is also difficult to justify, as the claimed statistical significance might have no practical impact. The need to answer questions like \textit{``Is method A better than B? Based on the experiments, how likely is A to be better? How high is the probability that A is better by more than 1\% ?''}, imposes the use of posterior probabilities. These are provided by the Bayesian methods described in \cite{edwards1963bayesian, dickey1973scientific, berger1987testing}.

This paper is organised in the following sections: Section \ref{rel_work} presents a review of the relevant literature which motivates the study performed in the current work and presents the assumptions upon which it is built. Section \ref{data} briefly describes the data used throughout the experiments. Section \ref{exps} provides an exhaustive description of the experiments conducted. Section \ref{results} presents and analyses the results and Section \ref{conclusions} concludes the work with ideas for further research efforts.

\section{Related Work} \label{rel_work}

%The work by Bouchaud \textit{et al.} \cite{bouchaud2009markets} offers an extensive and clear analysis of order-driven markets. Order book dynamics and determinants of the bid-ask spread, hence of trading costs, relevant to the present work are studied in the above-mentioned work.

The review by Bouchaud \textit{et al.} \cite{bouchaud2009markets} offers a thorough introduction to Limit Order Books, to which the interested reader is referred. As discussed in Section \ref{intro}, the growth of electronic trading has sparked the interest in Deep Learning applications to order-driven markets and Limit Order Books. The work by Kearns and Nevmyvaka \cite{kearns2013machine} presents an overview of Machine Learning and Reinforcement Learning applications to market microstructure data and tasks, including return prediction from Limit Order Book states. To our knowledge, the first attempt to produce an extensive analysis of Deep Learning-based methods for stock price prediction based upon the Limit Order Book was made by Tsantekidis \textit{et al.} \cite{tsantekidis2017forecasting}. In that paper, starting from a \textit{horse racing}-type comparison between classical Machine Learning approaches (e.g. Support Vector Machines) and more structured Deep Learning ones, they then considered the possibility to apply CNNs to detect anomalous events in the financial markets, and take profitable positions. In the last two years a few works applying a variety of Deep Learning models to LOB-based return prediction were published by the research group of Stephen Roberts, the first one, to the best of our knowledge, applied Bayesian (Deep Learning) Networks to Limit Order Book \cite{zhang2018bdlob}, followed by an augmentation to the labelling system as quantiles of returns and an adaptation of the modeling technique to this \cite{zhang2019extending}. The most recent work introduces the current state-of-the-art modeling architecture combining CNNs and LSTMs to delve deeper into the complex structure of the Limit Order Book. The work by Sirignano and Cont \cite{sirignano2019universal} provides a more theoretical approach, where it tests and compares multiple Deep Learning architectures for return forecasting based on order book data and shows how these models are able to capture and generalise to universal price formation mechanisms throughout assets.

The models used in the above works were originally defined in the literature from the field of Machine and Deep Learning and are summarised hereafter.

Multinomial Logistic Regression is used as a baseline for this work and consists in a linear combination of the inputs mapped through a \verb|logit| activation function, as defined in \cite{greene2003econometric}.
Feedforward Neural Networks (or Multilayer Perceptrons) are defined in \cite{haykin2004comprehensive} and constitute the general framework to represent non-linear function mappings between a set of input variables and a set of output variables. Recurrent Neural Networks (RNNs) \cite{lipton2015critical} are considered in the form of Long-Short Term Memory models (LSTMs) \cite{hochreiter1997long}. RNNs constitute an evolution of MLPs. They introduce the concept of sequentiality into the model including edges which span adjacent time steps. RNNs suffer from the issue of vanishing gradients when carrying on information for a large number of time steps.  LSTMs solve this issue by replacing nodes in the hidden layers with self-connected memory cells of unit edge weight which allow to carry on information without vanishing or exploding gradients. LSTMs hence owe their name to the ability to retain information through a long sequence. The addition of Attention mechanisms \cite{vaswani2017attention} to MLPs helps the model to focus more on relevant regions of the input data in order to make predictions. Self-Attention extends the parametric flexibility of global Attention Mechanisms by introducing an Attention mask that is no longer fixed, but a function of the input. The last kind of Deep Learning unit considered are Convolutional Neural Networks (CNNs), designed to process data with grid-like topology. These unit serve as feature extractors, thus learning feature representations of their inputs \cite{lecun1995convolutional}.

A considerable body of literature about comparison of different models has been produced, despite not being vastly applied by the Machine Learning community. The first attempts of formalisation were made by Ditterich \cite{dietterich1998approximate} and Salzberg \cite{salzberg1997comparing}, and refined by Nadau and Bengio \cite{nadeau2000inference} and Alpaydm \cite{alpaydm1999combined}. A comprehensive review of all these methods and of classical statistical tests for Machine Learning is presented in \cite{japkowicz2011evaluating}. A crucial point of view is provided by the work \cite{demvsar2008appropriateness}. More recently, starting from the work by Corani and Benavoli \cite{corani2015bayesian}, a Bayesian approach to statistical testing was proposed to replace classical approaches based on the null hypothesis. The proposed new ideas found a complete definition in \cite{benavoli2017time}.

\section{Dataset} \label{data}
All the experiments presented in this paper are based on the usage of the LOBSTER \cite{lobster_data} dataset, which provides a highly detailed, event-by-event description of all micro-scale market activities for each stock listed on the NASDAQ exchange. LOBSTER is one of the data providers featured in some major publications and journals in this field. LOB datasets are provided for each security in the NASDAQ. The dataset lists every market order arrival, limit order arrival and cancellation that occurs in the NASDAQ platform between 09:30 am – 04:00 pm on each trading day. Trading does not occur on weekends or public holidays, so these days are excluded from all the analyses performed. A tick size of $\theta = \$0.01$ is adopted. Depending on the type of the submitted order, orders can be executed at the lower cost equals of $\$0.005$. This is the case of hidden orders which, when revealed, appear at a price equal to the notional mid-price at the time of execution.
%We report below the data description provided in \cite{bouchaud_bonart_donier_gould_2018}. %``\textit{On the NASDAQ platform, each stock is traded into a separate LOB with a tick size of $\theta = \$0.01$. The platform enables traders to submit both visible and hidden limit orders. Visible limit orders obey standard price–time priority, while hidden limit orders have lower priority than all visible limit orders at the same price. The platform also allows traders to submit mid-price-pegged limit orders. These orders are hidden, but when executed, they appear with a price equal to the national mid-price (i.e. the mid-price calculated from the national best bid and offer) at their time of execution. Therefore, although the tick size is $\$0.01$, some orders are executed at a price ending with $\$0.005$. The employed dataset lists every market order arrival, limit order arrival, and cancellation that occurs on the NASDAQ platform during 09:30–16:00 each trading day. Trading does not occur on weekends or public holidays, so these days are excluded from all the performed analysis.}''

LOBSTER \cite{lobster_data} data are structured into two different files:

\begin{itemize}
    \item The \textit{message file} lists every market order arrival, limit order arrival and cancellation that occurs.
    
    \item The \textit{orderbook file} describes the market state (i.e. the total volume of buy or sell orders at each price) immediately after the corresponding event occurs.
    
\end{itemize}

Experiments described in the next few sections are performed only using the \textit{orderbook files}. The training dataset consists of Intel Corporation's (\verb|INTC|) LOB data from 04-02-2019 to 31-05-2019, corresponding to a total of 82 files, while the test dataset consists of Intel Corporation's LOB data from 03-06-2019 to 28-06-2019, obtained from 20 other files.
It is relevant to highlight that \verb|INTC| is representative of a large tick stock, these are stocks where the tick size is relative large compared to the price. These stocks present a range of specific characteristics in their LOB and trading dynamics and have been observed to be more predictable, through Deep Learning models, than small tick stocks. Hence, most of the market microstructure-related AI literature considers large tick stocks.
All the experiments presented in the current work are conducted on snapshots of the LOB with a depth (number of tick size-separated limit order levels per side of the Order Book) of $10$. This means that each row in the \textit{orderbook files} corresponds to a vector of length 40. Each row is structured as

\begin{equation}
[(p, v)^{a}_0, (p, v)^{b}_0, (p, v)^{a}_1, (p, v)^{b}_1, \, ... \, , (p, v)^{a}_{10}, (p, v)^{b}_{10}],
\label{lob_state_def}
\end{equation}

%\begin{equation}
%[p^{a}_0, v^{a}_0, p^{b}_0, v^{b}_0, p^{a}_1, v^{a}_1, p^{b}_1, v^{b}_1, \, ... %\, , p^{a}_{10}, v^{a}_{10}, p^{b}_{10}, v^{b}_{10}],
%\label{lob_state_def}
%\end{equation}

where $(p, v)$ represents the price level and corresponding liquidity tuple, $\{a,b\}$ distinguish ask and bid levels progressively further away from the best ask and best bid.

\section{Methods} \label{exps}

\subsection{Price change horizons} \label{horizon_def}

Price log-returns for the target labels are defined at three distinct time horizons $H_{\Delta \tau}$. In order to account for price volatility and discount long periods of stable and noisy order flow, the time delay between the LOB observation (input) and the target label return $\Delta \tau$ is defined as follows.

Given a series of mid-prices at consecutive ticks

\begin{equation}
    p_{m, 0}, p_{m, 1}, \, ... \, ,  p_{m, n},
\end{equation}

the mid-price is defined as the mean between the best bid and best ask price. The series of log-returns is

\begin{equation}
    r_{m, 0}, r_{m, 1}, \, ... \, ,  r_{m, n-1},
\end{equation}

where 

\begin{equation}
    r_{m, 0} = \log p_{m, 1} - \log p_{m, 0}.
\end{equation}

The number of non-zero log-returns in the series is hence counted as:

\begin{equation}
    \Delta \tau = \sum_{k = 0}^{n-1} \Theta (|r_{m, k}|),
\end{equation}

where $\Theta$ is the Heaviside step function defined below

\begin{equation}
    \Theta(x) = 
         \begin{cases}
           1 & \text{if }\ x > 0 \\
           0 & \text{if }\ x \leq 0.
         \end{cases}
\end{equation} 

\subsection{Data preprocessing and labelling} \label{data_prep}

The data described in Section \ref{data} are preprocessed as follows:

\begin{itemize}

%The target labels for the prediction task aim to categorise the return at three different time horizons $H_{\Delta \tau}$. In order to perform the current mapping from their numerical values into discrete classes, the following quantile levels $(0., 0.25, 0.75, 1.)$ are hence computed on the returns distribution of the training set. These quantiles are mapped into classes as reported in Figure (...).

%%% FIGURE
    %\item For each tick $t$ (i.e. each row of the corresponding dataset) the quantile bin edges ($0., 0.25, 0.75, 1.$) of the mid-price at $H_{\Delta \tau} | \Delta \tau \in \{ 10, 50, 100 \}$, are computed as per the definition in Section \ref{horizon_def}.Quantiles are computed on the whole training set and then applied to the test set. Quantile ranges encode the return sign and magnitude into three classes (\textit{down}, \textit{flat}, \textit{up}) at each of the above-mentioned horizons $H_{\Delta \tau}$ and they are mapped as follows:
    
    \item The target labels for the prediction task aim to categorise the return at three different time horizons $H_{\Delta \tau} | \Delta \tau \in \{ 10, 50, 100 \}$. In order to perform the mapping from continuous variables into discrete classes, the following quantile levels $(0., 0.25, 0.75, 1.)$ are computed on the returns distribution of the training set and then applied to the test set. These quantiles are mapped onto classes, denoted with $(q_{-1}, q_0, q_{+1})$ as reported in Figure \ref{Quantiles_representation}.
    
    \begin{figure}[H]
        \centering
        \includegraphics[scale=0.22]{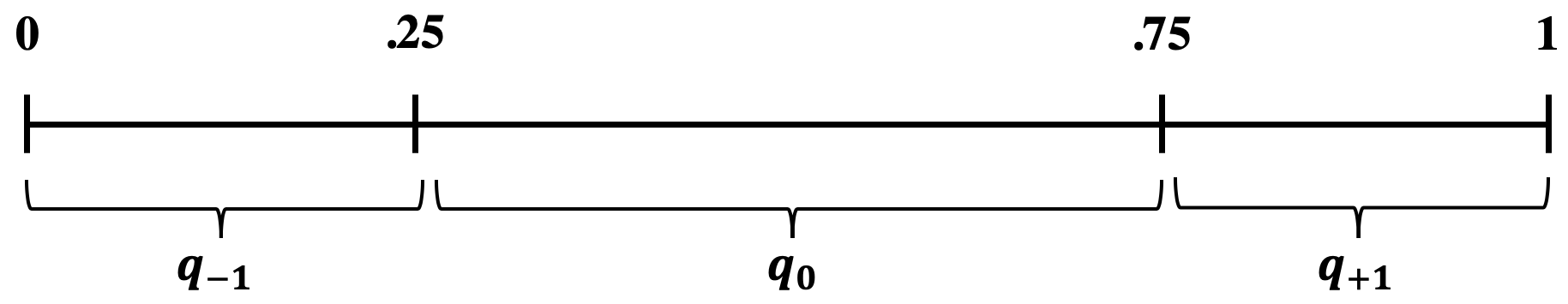}
        \caption{Visual representation of the mapping between quantiles and corresponding classes. Quantiles' edges (i.e. $(0., 0.25, 0.75, 1.)$) define three different intervals. Each specific class (i.e. $q_{-1}$, $q_{0}$, $q_{+1}$) corresponds to a specific interval.}
        \label{Quantiles_representation}
    \end{figure}
    
    \item The training set input data (LOB states) are scaled within a $(0, 1)$ interval with the \verb|min-max| scaling algorithm \cite{scikit-learn}. The scaler's training phase is conducted by chunks to optimise the computational effort. The trained scaler is then applied to the test data.
\end{itemize}

\begin{figure}[H]
    \centering
    \includegraphics[width=1\linewidth]{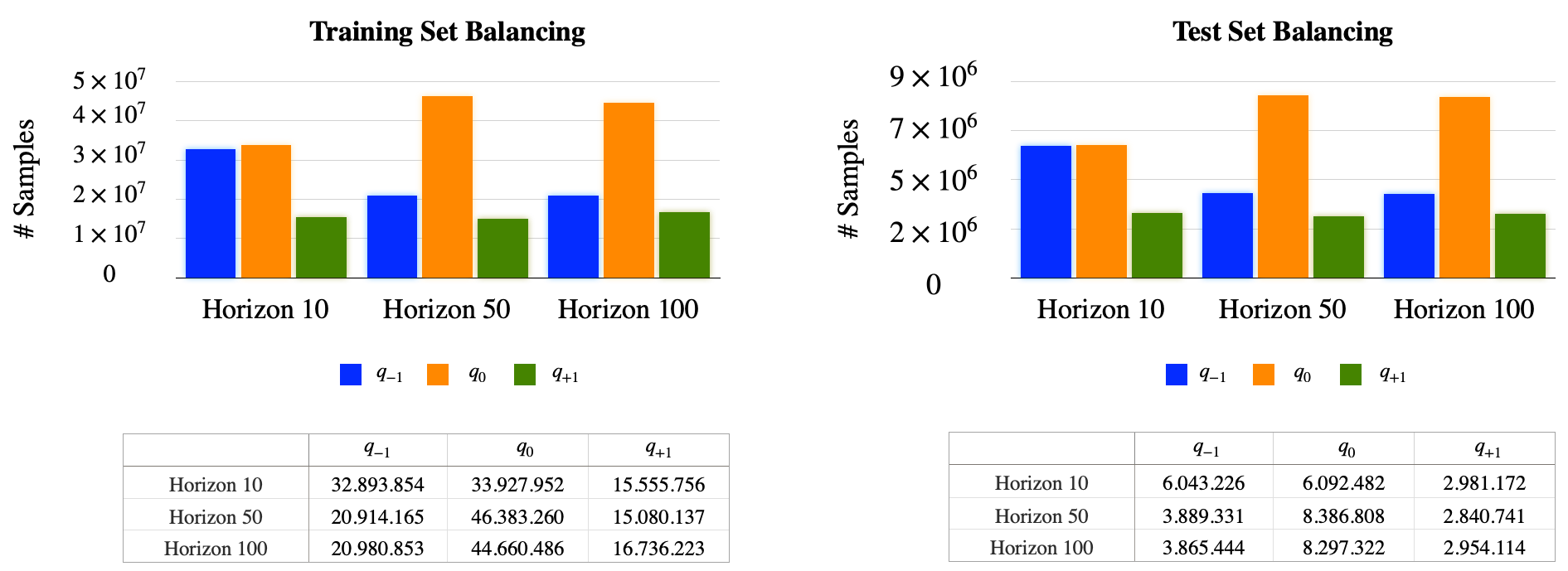}
    \caption{Training and test set quantile ($q_{-1}$, $q_{0}$, $q_{+1}$) distributions per horizon $H_{\Delta \tau} |\Delta \tau \in \{10, 50, 100\}$ at the end of the preprocessing and labelling phase. Tables' entries, for both the training and the test set, report the exact number of samples per horizon, for each considered quantile.}
    \label{Training_Set_Balancing}
\end{figure}

Figure \ref{Training_Set_Balancing} reports the training and test set quantile distributions per horizon $H_{\Delta \tau}$. %at the end of the preprocessing phase.

It is possible to notice moderately balanced classes for both plots in Figure \ref{Training_Set_Balancing}. Indeed, all classes lie within the same order of magnitude ($10^7$) for all horizons  $H_{\Delta \tau} |\Delta \tau \in \{10, 50, 100\}$ with the $q_0$ class being the most represented and $q_{+1}$ the least.

\subsection{Random Model} \label{random_model}
The benchmark null model for this work is a generic random model, which does not handle any dynamics. For each sample in the test set and each horizon $H_{\Delta \tau}$, the quantile label $q_r$ is sampled from the uniform distribution over $r \in \{-1, 0, 1\}$. The SciPy \cite{2020SciPy-NMeth} \verb|randint| generator is used for this task.

\subsection{Naive Model} \label{naive_model}
In order to ensure a fair comparison, the benchmark naive model for this work improves upon the model in Section \ref{random_model} by overfitting to the most present class in the training set (See Figure \ref{Training_Set_Balancing}). For each sample in the test set and each horizon $H_{\Delta \tau}$, the predicted quantile label $q_r$ is hence always $q_0$.

\subsection{Logistic Regression}
The baseline model is represented by the multinomial Logistic Regression which, as the Random Model, does not explictly model any dynamics in the data. Like binary Logistic Regression, the multinomial one adopts maximum likelihood estimation to evaluate the probability of categorical membership. It is also known as Softmax Regression and can be interpreted as a classical ANN, as per the definition in Table \ref{tab:Logistic_Regression}, with the input layer directly connected to the output layer with a \textit{softmax} activation function:

\begin{equation}
    \text{softmax}(x_{i}) = \frac{e^{x_i}}{\sum_j e^{x_j}}. 
    \label{softmax_function}
\end{equation}

The input is represented by the ten most recent LOB states as per the definition is Equation \ref{lob_state_def} in Section \ref{data}. Also this model is not able to handle any specific dynamics. The Scikit-Learn \cite{scikit-learn} implementation is used and, in order to guarantee a fair comparison with the Deep Learning models in the next sections, the following parameters are set as follows:

\begin{itemize}
    \item \textit{max\_iter} (i.e. maximum number of iterations taken for the solvers to converge) = 20.
    \item \textit{tol} (i.e. the tolerance for stopping criteria) = $e^{-1}$.
    \item \textit{solver} (i.e. the algorithm to use in the optimization problem) = \verb|sag| with the default $L2$ penalty.
\end{itemize}

\begin{table}[H]
\centering
\caption{Multinomial Logistic Regression architectural scheme.}
\label{tab:Logistic_Regression}
    \begin{adjustbox}{scale=0.8,center}
        \begin{tabular}{@{}c@{}}
        \toprule
        \textbf{Logistic regression} \\ \midrule
        \begin{tabular}[c]{@{}c@{}}Input @ {[}$1 \times 400${]}\\ \\ Dense @ 3 Units (activation \verb|softmax|)\end{tabular} \\ \bottomrule
        \end{tabular}%
    \end{adjustbox}
\end{table}

\subsection{Multilayer Perceptron}
The first Deep Learning model is a generic Multilayer Perceptron (MLP), which does not explicitly model temporal or spatial properties of the input data, but has the ability to model not explicitly defined dimensions through its hidden layers. Similarly to the two previously mentioned models, it does not explicitly handle any specific dimension. The MLP can be considered the most general form of universal approximator and it represents the ideal model to confirm or reject any hypothesis about the presence of a specific leading dimension in LOBs.

In order to allow a fair comparison with the other sequence-based models, the input for the MLP is represented by a $40 \cdot 10 = 400$ vector containing the ten most recent LOB states (see Equation \ref{lob_state_def}) concatenated in a flattened shape. The MLP model is architecturally defined in Table \ref{tab:Experiment_6_Multilayer_Perceptron}.

\begin{table}[H]
\centering
\caption{Multilayer Perceptron architectural scheme.}
\label{tab:Experiment_6_Multilayer_Perceptron}
    \begin{adjustbox}{scale=0.8,center}
        \begin{tabular}{@{}c@{}}
        \toprule
        \textbf{Multilayer Perceptron} \\ \midrule
        \begin{tabular}[c]{@{}c@{}}Input @ {[}$10 \times 40${]}\\ \\ Dense @ 512 Units \\ Dense @ 1024 Units\\ Dense @ 1024 Units\\ Dense @ 64 Units\\ \\ Dense @ 3 (activation \verb|softmax|)\end{tabular} \\ \bottomrule
        \end{tabular}%
    \end{adjustbox}
\end{table}

\subsection{Shallow LSTM}
\label{shallow_LSTM_def}
In order to explicitly handle temporal dynamics of the system, a shallow LSTM model is tested \cite{hochreiter1997long}. The LSTM architecture explicitly models temporal and sequential dynamics in the data, hence providing insight on the temporal dimension of the data. As all the other RNN models, the structure of LSTMs enables the network to capture the temporal dynamics performing sequential predictions. The current state directly depends on the previous ones, meaning that the hidden states represent the memory of the network. Differently from classic RNN models, LSTMs are explicitly designed to overcome the vanishing gradient problem as well as capture the effect of long-term dependencies. The input is here represented by a $[10 \times 40]$ matrix, where 10 is the number of consecutive history ticks and 40 is the shape of the LOB defined in Equation \ref{lob_state_def}. The LSTM layer consists of 20 units with a $\tanh$ activation function:

\begin{equation}
   \tanh(x) = \frac{e^x-e^{-x}}{e^x+e^{-x}}
   \label{tanh_activation}
\end{equation}

It is observed that the addition of LSTM units beyond the chosen level does not yield statistically significant performance improvements. Hence, the chosen number of LSTM units can be considered optimal and the least computationally costly. The model is architecturally defined in Table \ref{tab:Experiment_2_Shallow_LSTM}.

\begin{table}[H]
\centering
\caption{Shallow LSTM architectural scheme.}
\label{tab:Experiment_2_Shallow_LSTM}
    \begin{adjustbox}{scale=0.8,center}
        \begin{tabular}{@{}c@{}}
        \toprule
        \textbf{Shallow LSTM} \\ \midrule
        \begin{tabular}[c]{@{}c@{}}Input @ {[}$10 \times 40${]}\\ \\ LSTM @ 20 Units\\ \\ Dense @ 3 Units (activation \verb|softmax|)\end{tabular} \\ \bottomrule
        \end{tabular}%
    \end{adjustbox}
\end{table}

\subsection{Self-Attention LSTM}
\label{Self_Attention_def}
As a point of contact between architectures which model temporal dynamics (LSTMs) and spatial modeling ones (CNNs), the LSTM described in Section \ref{shallow_LSTM_def} is enhanced by the introduction of a Self-Attention module \cite{vaswani2017attention}. By default, the Attention layer statically considers the whole context while computing the relevance of individual entries. Differently from what described in Section \ref{CNN-LSTM}, the input is not subject to any spatial transformation (e.g. convolutions). This difference implies a static nature of the detected behaviours over multiple timescales. The Self-Attention LSTM is architecturally defined in Table \ref{tab:Experiment_7_Self-attention}.

\begin{table}[H]
\centering
\caption{Self-Attention LSTM architectural scheme.}
\label{tab:Experiment_7_Self-attention}
    \begin{adjustbox}{scale=0.8,center}
        \begin{tabular}{@{}c@{}}
        \toprule
        \textbf{Self-attention LSTM} \\ \midrule
        \begin{tabular}[c]{@{}c@{}}Input @ {[}$10 \times 40${]}\\ \\ LSTM @ 40 Units\\ \\ Self-Attention Module\\ \\ Dense @ 3 (activation \verb|softmax|)\end{tabular} \\ \bottomrule
        \end{tabular}%
    \end{adjustbox}
\end{table}

The input to this model is represented by a $[10 \times 40]$ matrix, where 10 is the number of consecutive history ticks and 40 is the shape of the LOB defined in Equation \ref{lob_state_def}.

\subsection{CNN-LSTM}
\label{CNN-LSTM}
The nature of the Deep Learning architectures in Sections \ref{shallow_LSTM_def}, \ref{Self_Attention_def} mainly focuses on modeling the temporal dimension of the inputs. Recent developments \cite{DeepLOB, sirignano2019deep} highlight the potential of the spatial dimension in LOB-based forecasting as a structural module allowing to capture dynamic behaviours over multiple timescales. In order to study the effectiveness of such augmentation, the architecture described in \cite{DeepLOB, briola_antonio_and_turiel_jeremy_david_2020_4104275} is reproduced as in Table \ref{tab:Experiment_5_CNN_LSTM}, and adapted to the application domain described in the current work. The input is represented by a $[10 \times 40]$ matrix, where 10 is the number of consecutive history ticks and 40 is the shape of the LOB defined in Equation \ref{lob_state_def}. This model represents the state-of-the-art in terms of prediction potential at the time of writing. 

\begin{table}[H]
\centering
\caption{CNN-LSTM architectural scheme.}
\label{tab:Experiment_5_CNN_LSTM}
    \begin{adjustbox}{scale=0.8,center}
        \begin{tabular}{@{}c@{}}
        \toprule
        \textbf{CNN-LSTM} \\ \midrule
        \begin{tabular}[c]{@{}c@{}}Input @ {[}$10 \times 40${]}\\ \\ Conv\\ $1 \times 2$ @ 16 (stride = $1 \times 2$)\\ $4 \times 1$ @ 16\\ $4 \times 1$ @ 16\\ \\ $1 \times 2$ @ 16 (stride = $1 \times 2$)\\ $4 \times 1$ @ 16\\ $4 \times 1$ @ 16\\ \\ $1 \times 10$ @ 16\\ $4 \times 1$ @16\\ $4 \times 1$ @ 16\\ \\ Inception @ 32\\ \\ LSTM @ 64 Units\\ \\ \\ Dense @ 3 (activation \verb|softmax|)\end{tabular} \\ \bottomrule
        \end{tabular}%
    \end{adjustbox}
\end{table}

\subsection{Training pipeline} \label{train_pipe}

%A range of Deep Learning models able to handle different dimensions is considered. Table \ref{tab:Methods_Table} provides a summary of the models. 
For each of the Deep Learning models, the following training (and testing, see Section \ref{test_pipe}) procedure is applied:

\begin{itemize}
    \item Training batches of 1024 samples are produced, with each sample made of 10 consecutive LOB states. LOB states are defined in Section \ref{data} and in Equation \ref{lob_state_def}.
    \item For each training epoch $1.6 \times 10^4$ batches are randomly chosen. This number of batches is selected to consider a total number of training samples equivalent to one month ($\sim 1,7 \times 10^7$ samples). This sampling procedure ensures a good coverage of the entire dataset and allows to operate with a reduced amount of computational resources.
    \item Class labels are converted to their one-hot representation. 
    \item The selected optimizer is \verb|Adam| \cite{kingma2014adam}. Its Keras \cite{keras_description} implementation is chosen and default values for its hyperparameters are kept (\verb|lr|$ = 0.001$). The \verb|categorical crossentropy| loss function is chosen due to its suitability for multi-class classification tasks \cite{francois2017deep}.
    \item From  manual hyperparameter exploration it is observed that 30 training epochs are optimal when accounting for constraints on computational resources. It has been empirically observed that slight variations do not produce any significant improvement.
\end{itemize}

\subsection{Test pipeline and performance metrics} \label{test_pipe}

At the end of the training phase, the inducer for each model is queried on the test set as follows:

\begin{itemize}
    \item The Keras \cite{keras_description} Time Series Generator is used to rearrange the test set creating batches of $5\times10^5$ test samples. Each sample is made of 10 consecutive states (Equation \ref{lob_state_def}). 
    %A total number of 31 independent test sets are created according to this definition. This number guarantees statistical significance for the considered problem \cite{lenth2001some}.
    \item For each model and test fold, balanced Accuracy \cite{mosley2013balanced, kelleher2015fundamentals, guyon2015design}, weighted Precision, weighted Recall and weighted F-score are computed. These metric are weighted in order to correct for class imbalance and obtain unbiased indicators. The following individual class metrics are considered too: Precision, Recall and F-measure. Two multi-class correlation metrics between labels are also computed: Matthews Correlation Coefficient (MCC) \cite{gorodkin2004comparing} and Cohen’s Kappa \cite{cohen1960coefficient, artstein2008inter}.
    \item Performance metrics for each test fold are statistically compared through the Bayesian correlated t-test \cite{corani2015bayesian, benavoli2017time}. One should note that the \textit{region of practical equivalence} (rope) determining the negligible difference between performance metrics in different models, is arbitrarily set to a sensible $3\%$, due to the lack of examples in the literature.
    %Performance metrics for each one of the 31 test folds are statistically compared through the Bayesian correlated t-test \cite{corani2015bayesian, benavoli2017time}. One should note that the \textit{region of practical equivalence} (rope) determining the negligible difference between performance metrics in different models, is arbitrarily set to a sensible $3\%$, due to the lack of examples in the literature.
    
\end{itemize}

\begin{table}[H]
\centering
\caption{A summary of the Deep Learning models and related dynamics.}
\label{tab:Methods_Table}
\resizebox{8cm}{!}{%
\begin{tabular}{@{}cc@{}}
\toprule
\textbf{Model} & \textbf{Dimension} \\ \midrule
Random Model \& Naive Model & None \\
Multinomial Logistic Regression & None \\
Multilayer Perceptron & Not explicitly defined \\
Shallow LSTM & Temporal \\
Self-Attention LSTM & Temporal + Spatial (static) \\
CNN-LSTM & Temporal + Spatial (dynamic) \\ \bottomrule
\end{tabular}%
}
\end{table}

\section{Results} \label{results}

Multinomial Logistic Regressions, MLPs, LTSMs, LSTMs with Attention and CNN-LSTMs are trained to predict the return quantile $q$ at different horizons $H_{\Delta \tau} |\Delta \tau \in \{10, 50, 100\}$. The dataset used for all models is defined in Section \ref{data} and the metrics used to evaluate and compare out of sample model performances are introduced in Section \ref{test_pipe}.
%The models are ranked based on those metrics and they are clustered as visualised in Figure \ref{fig:performances_clusters} through the Bayesian correlated t-test defined in Section \ref{test_pipe}. A reflection on the underlying dimensions of the Limit Order Book, as per the introductory discussion in Sections \ref{intro}, \ref{rel_work}, concludes the section.

\begin{figure}[H]
    \centering
    \includegraphics[scale=0.24]{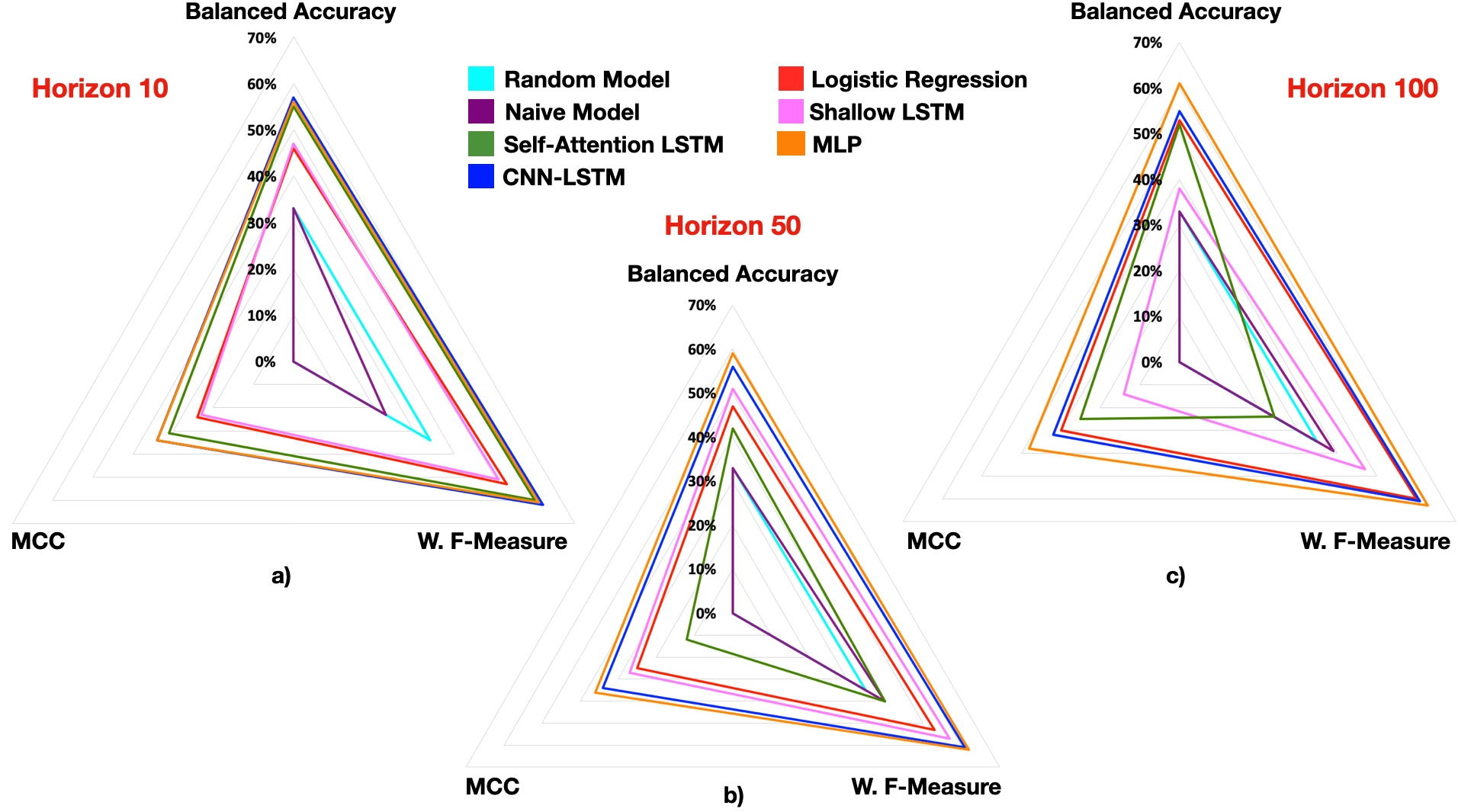}
    \caption{Radar plot comparison of the model described in Section \ref{exps}. For each model (different color triangles) a closer node to the outer boundary of the scale indicates a higher value of the corresponding metric. The plot hence provide an intuitive understanding of model superiority and whether this is consistent throughout metrics. From this figure it is evident that the MLP performs best, according to the three measures considered.}
    \label{fig:radar}
\end{figure}

Out of sample performance metrics are reported in Table \ref{tab:results_performance_metrics_horizon_10_50_100} and visualised in Figure \ref{fig:radar}. 
Models are clustered into three groups, based on their performance metrics. Specifically, in each one of these clusters it is possible to locate models that perform statistically equivalent throughout horizons $H_{\Delta \tau}$, based on the MCC and weighted F-measure metrics described in Section \ref{train_pipe}. A representation of model clustering and ordering is presented in Figure \ref{fig:performances_clusters}, Appendix \ref{appendix: appendixA}.

Similarities between models' performances are tested by means of the Bayesian Correlated t-test. This allowed to assign the models to the relative clusters or intersections of those as per the representation in Figure \ref{fig:performances_clusters}, Appendix \ref{appendix: appendixA}.
It is important to clarify that the McNemar Test is not applied to compare models in this work, due to the belief that the Bayesian Correlated t-test is more appropriate for the present experimental setup. Further, from theoretical considerations, one expects to obtain analogous results to the presented Bayesian test. Future work shall include additional tests.
%Analogous results to the ones obtained through the Bayesian test are expected.

\begin{table}[H]
\centering
\caption{Performance metrics for horizons $H_{\Delta \tau}$ computed on the test folds. The column labels \textbf{H10}, \textbf{H50, \textbf{H100}} refer to $\mathbf{H_{\Delta \tau} | \Delta \tau = 10}$, $\mathbf{H_{\Delta \tau} | \Delta \tau = 50}$, $\mathbf{H_{\Delta \tau} | \Delta \tau = 100}$, respectively.}
\label{tab:results_performance_metrics_horizon_10_50_100}
\resizebox{\textwidth}{!}{%
\begin{tabular}{@{}
>{\columncolor[HTML]{FFFFFF}}c 
>{\columncolor[HTML]{FFFFFF}}c 
>{\columncolor[HTML]{FFFFFF}}c 
>{\columncolor[HTML]{FFFFFF}}c 
>{\columncolor[HTML]{FFFFFF}}c 
>{\columncolor[HTML]{FFFFFF}}c 
>{\columncolor[HTML]{FFFFFF}}c 
>{\columncolor[HTML]{FFFFFF}}c 
>{\columncolor[HTML]{FFFFFF}}c 
>{\columncolor[HTML]{FFFFFF}}c 
>{\columncolor[HTML]{FFFFFF}}c 
>{\columncolor[HTML]{FFFFFF}}c 
>{\columncolor[HTML]{FFFFFF}}c 
>{\columncolor[HTML]{FFFFFF}}c 
>{\columncolor[HTML]{FFFFFF}}c 
>{\columncolor[HTML]{FFFFFF}}c 
>{\columncolor[HTML]{FFFFFF}}c 
>{\columncolor[HTML]{FFFFFF}}c 
>{\columncolor[HTML]{FFFFFF}}c
>{\columncolor[HTML]{FFFFFF}}c 
>{\columncolor[HTML]{FFFFFF}}c 
>{\columncolor[HTML]{FFFFFF}}c @{}}
\toprule
 & \multicolumn{3}{c}{\cellcolor[HTML]{FFFFFF}\textbf{Random Model}} & \multicolumn{3}{c}{\cellcolor[HTML]{FFFFFF}\textbf{Naive Model}} & \multicolumn{3}{c}{\cellcolor[HTML]{FFFFFF}\textbf{Logistic Regression}} & \multicolumn{3}{c}{\cellcolor[HTML]{FFFFFF}\textbf{Shallow LSTM}} & \multicolumn{3}{c}{\cellcolor[HTML]{FFFFFF}\textbf{Self-Attention LSTM}} & \multicolumn{3}{c}{\cellcolor[HTML]{FFFFFF}\textbf{CNN-LSTM}} & \multicolumn{3}{c}{\cellcolor[HTML]{FFFFFF}\textbf{Multilayer Perceptron}} \\ \midrule
 & \textbf{H10} & \textbf{H50} & \textbf{H100} &\textbf{H10} & \textbf{H50} & \textbf{H100} &  \textbf{H10} & \textbf{H50} & \textbf{H100} & \textbf{H10} & \textbf{H50} & \textbf{H100} & \textbf{H10} & \textbf{H50} & \textbf{H100} & \textbf{H10} & \textbf{H50} & \textbf{H100} & \textbf{H10} & \textbf{H50} & \textbf{H100} \\ \midrule
\textbf{Balanced Accuracy} & 0.33 & 0.33 & 0.33 & 0.33 & 0.33 & 0.33 & 0.46 & 0.47 & 0.53 & 0.47 & 0.51 & 0.38 & 0.55 & 0.42 & 0.52 & 0.57 & 0.56 & 0.55 & 0.56 & 0.59 & 0.61 \\
\textbf{Weighted Precision} & 0.41 & 0.41 & 0.41 & 0.16 & 0.30 & 0.30 & 0.54 & 0.56 & 0.62 & 0.58 & 0.57 & 0.65 & 0.61 & 0.50 & 0.47 & 0.62 & 0.61 & 0.61 & 0.62 & 0.62 & 0.63 \\
\textbf{Weighted Recall} & 0.33 & 0.33 & 0.33 & 0.40 & 0.55 & 0.54 & 0.59 & 0.59 & 0.61 & 0.58 & 0.58 & 0.57 & 0.61 & 0.45 & 0.34 & 0.62 & 0.62 & 0.62 & 0.62 & 0.63 & 0.63 \\
\textbf{Weighted F-Measure} & 0.34 & 0.35 & 0.35 & 0.23 & 0.40 & 0.39 & 0.53 & 0.53 & 0.60 & 0.51 & 0.57 & 0.47 & 0.60 & 0.40 & 0.24 & 0.62 & 0.61 & 0.61 & 0.61 & 0.62 & 0.63 \\ \midrule
\textbf{Precision quantile {[}0, 0.25{]}} & 0.26 & 0.26 & 0.26 & 0 & 0 & 0 & 0.57 & 0.57 & 0.57 & 0.57 & 0.56 & 0.58 & 0.60 & 0.37 & 0.55 & 0.59 & 0.59 & 0.59 & 0.59 & 0.59 & 0.59 \\
\textbf{Precision quantile {[}0.25, 0.75{]}} & 0.55 & 0.55 & 0.55 & 0.40 & 0.55 & 0.54 & 0.59 & 0.60 & 0.63 & 0.59 & 0.62 & 0.56 & 0.62 & 0.55 & 0.52 & 0.65 & 0.64 & 0.63 & 0.64 & 0.66 & 0.67 \\
\textbf{Precision quantile {[}0.75, 1{]}} & 0.20 & 0.20 & 0.20 & 0 & 0 & 0 & 0.31 & 0.38 & 0.58 & 0.57 & 0.43 & 0.97 & 0.57 & 0.52 & 0.26 & 0.57 & 0.57 & 0.57 & 0.59 & 0.58 & 0.57 \\ \midrule
\textbf{Recall quantile {[}0, 0.25{]}} & 0.33 & 0.33 & 0.33 & 0 & 0 & 0 & 0.55 & 0.59 & 0.59 & 0.06 & 0.62 & 0.22 & 0.35 & 0.81 & 0.62 & 0.54 & 0.55 & 0.53 & 0.60 & 0.59 & 0.60 \\
\textbf{Recall quantile {[}0.25, 0.75{]}} & 0.33 & 0.33 & 0.33 & 1 & 1 & 1 & 0.82 & 0.81 & 0.76 & 0.85 & 0.69 & 0.93 & 0.76 & 0.44 & 0.01 & 0.71 & 0.72 & 0.74 & 0.74 & 0.70 & 0.68 \\
\textbf{Recall quantile {[}0.75, 1{]}} & 0.33 & 0.33 & 0.33 & 0 & 0 & 0 & 0 & 0 & 0.23 & 0.51 & 0.23 & 0 & 0.53 & 0.004 & 0.92 & 0.46 & 0.42 & 0.39 & 0.34 & 0.46 & 0.54 \\ \midrule
\textbf{F-Measure quantile {[}0, 0.25{]}} & 0.29 & 0.29 & 0.29 & 0 & 0 & 0 & 0.56 & 0.57 & 0.58 & 0.11 & 0.59 & 0.31 & 0.44 & 0.503 & 0.58 & 0.57 & 0.57 & 0.56 & 0.59 & 0.59 & 0.59 \\
\textbf{F-Measure quantile {[}0.25, 0.75{]}} & 0.42 & 0.42 & 0.42 & 0.57 & 0.71 & 0.70 & 0.70 & 0.70 & 0.68 & 0.69 & 0.65 & 0.70 & 0.68 & 0.489 & 0.02 & 0.68 & 0.68 & 0.68 & 0.69 & 0.68 & 0.67 \\
\textbf{F-Measure quantile {[}0.75, 1{]}} & 0.25 & 0.25 & 0.25 & 0 & 0 & 0 & 0 & 0 & 0.33 & 0.54 & 0.30 & 0 & 0.55 & 0.009 & 0.40 & 0.51 & 0.48 & 0.46 & 0.43 & 0.51 & 0.56 \\ \midrule
\textbf{MCC} & 0 & 0 & 0 & 0 & 0 & 0 & 0.24 & 0.25 & 0.30 & 0.23 & 0.27 & 0.14 & 0.31 & 0.120 & 0.25 & 0.34 & 0.34 & 0.32 & 0.34 & 0.36 & 0.38 \\
\textbf{Cohen's Kappa} & 0 & 0 & 0 & 0 & 0 & 0 & 0.21 & 0.22 & 0.30 & 0.20 & 0.27 & 0.10 & 0.30 & 0.105 & 0.16 & 0.34 & 0.33 & 0.32 & 0.33 & 0.36 & 0.38 \\ \bottomrule
\end{tabular}
}
\end{table}

\section{Discussion}

This section builds upon the results in Section \ref{results} and delves deeper into the analysis of model similarities and individual dimension-based model performances. A better Deep Learning-based understanding of LOB dynamics and modeling applications emerges from this.

Delving deeper into the analysis of model performances, the plots for the different prediction horizons in Figure \ref{fig:radar} show how the MLP outperforms the CNN-LSTM throughout metrics and horizons (or performs analogously). A second group of models is represented by Logistic Regression and Shallow LSTM for horizons $H_{\Delta \tau} |\Delta \tau \in \{10, 50\}$. The simplicity and stability of the Logistic Regression model benefits performance in $H_{\Delta \tau} | \tau =100$ where it performs analogously to the CNN-LSTM model. This is due to the CNN-LSTM model showing worsened performance, with respect to the MLP, for the longest horizon. A similar statement on depleted performance applies to the Shallow LSTM at this horizon. The Self-Attention LSTM performs well at $H_{\Delta \tau} | \tau =10$, perhaps benefiting from the collective overview in the noisy shorter horizon, but sees significantly worse and unstable performance for the longer horizons. All stably performing models perform well above the two baseline models (Random and Naive), hence showing the ability of the model to learn and the possibility to extract information from the complex and highly stochastic setting which characterises the LOB.
It is important to note that the MLP's performance slightly increases with horizon length, showing how meaningful features are being extracted such that the model benefits from longer-term trajectories less affected by HFT short-term price noise.

We now analyse the model clusters in Figure \ref{fig:performances_clusters}, Appendix \ref{appendix: appendixA}. The clusters are formed using the Bayesian Correlated t-test described by Benavoli \textit{et al.} \cite{benavoli2017time}. Models are considered statistically equivalent and hence assigned to the same cluster on the basis of their MCC and weighted F-measure metrics.

The first cluster of models %in Figure \ref{fig:performances_clusters}, Appendix \ref{appendix: appendixA} 
is characterised by the Random Model, and it also comprises Logistic Regression and Self-Attention LSTM at intersections with other groups. Looking at Table \ref{tab:results_performance_metrics_horizon_10_50_100}, low values for weighted Precision and weighted Recall are observed for the Random Model. This means that the system, for each horizon $H_{\Delta \tau}$, yields balanced predictions throughout classes and by chance a few of them are correct. Most of the correctly classified labels belong to the central quantile $q_0$, as shown by the higher values of its class-specific Precision. The second most correctly predicted quantile is the lower one $q_{-1}$, while the worst overall performance is associated with the upper quantile $q_{+1}$. Given that these predictions are picked randomly from an uniform distribution, the obtained results reflect the test set class distribution. The  MCC  and  the  Cohen’s Kappa are both equal to zero, thereby confirming that the model performs in a  random fashion. Analogous considerations can be made for the Naive Model. \\

It results difficult to assign multinomial Logistic Regression to a cluster, as results in Table \ref{tab:results_performance_metrics_horizon_10_50_100} show that the model is solid throughout horizons $H_{\Delta \tau}$ but lacks the ability to produce complex features which could improve its performance, due to the absence of non-linearities and hidden layers. Because of its behaviour, it is placed at the overlap between the first two clusters. Looking at Tables \ref{tab:weighted_mcc_horizon_10} and \ref{tab:weighted_mcc_horizon_50}, it is possible to note how the multinomial Logistic Regression consistently outperforms the Random Model. For longer horizons (i.e. $H_{\Delta \tau} |\Delta \tau \in \{50, 100\}$), Logistic Regression 
outperforms more structured Deep Learning models designed to handle specific dynamics. Despite this result, the model is systematically unable to decode the signal related to the upper quantile, perhaps due to the slight class imbalance in the training set. \\

The second cluster of models is represented by the shallow LSTM model. In this case, weighted Precision and weighted Recall values for all three horizons $H_{\Delta \tau}$ are higher than for Logistic Regression and the Random Model. This means that the proposed model is able to correctly predict a considerably high number of samples from different classes, when compared to the previously considered architectures. Similarly to what described in the previous paragraph, the model's predictions are well balanced between quantiles, for horizons $H_{\Delta \tau} | \Delta \tau \in \{ 10, 50 \}$. This observation is confirmed by higher values for the MCC and Cohen's Kappa for $H_{\Delta \tau} | \Delta \tau \in \{ 10, 50 \}$, which indicate that an increasing number of predictions match the ground-truth. Performance metrics collapse for the longer horizon $H_{\Delta \tau} | \Delta \tau = 100$ in the upper quantile $q_{+1}$.\\

Results achieved by the Attention-LSTM make assigning the model to one of the three clusters
%in Figure \ref{fig:performances_clusters}, Appendix \ref{appendix: appendixA}
extremely difficult. As stated for class specific performances in the shallow LSTM model, there is a strong relation between metrics and the considered horizon $H_{\Delta \tau}$. For $H_{\Delta \tau} | \Delta \tau = 10$, higher values of weighted Precision are accompanied by higher values of weighted Recall. These imply that the model is able to correctly predict a considerable number of samples from different classes. The only exception is represented by the lower quantile $q_{-1}$ which has a lower class-specific weighted Recall. Looking at Table \ref{tab:results_performance_metrics_horizon_10_50_100}, it is possible to note higher MCC and Cohen's Kappa values for the considered model, suggesting a relatively structured agreement between predictions and the ground truth, as described in Table \ref{tab:weighted_mcc_horizon_10}. This makes the Attention-LSTM model statistically equivalent to the state-of-the-art models (namely the CNN-LSTM and MLP).
Our analysis changes significantly when considering $H_{\Delta \tau} | \Delta \tau = 50$. A decrease of more than $10\%$ in all performance metrics is observed. The greatest impact is on the upper quantile where the model is less capable to perform correct predictions. All these considerations strongly impact the Matthews Correlation Coefficient which has a value $50 \%$ lower than the one for $H_{\Delta \tau} |\Delta \tau = 10$. The last analysis concerns the results obtained for $H_{\Delta \tau} | \Delta \tau = 100$. For this horizon, the model yields more balanced performances in terms of correctly predicted samples for the extreme quantiles $q_{-1,+1}$, while the greatest impact in terms of performance is on $q_0$. It is relevant to highlight the high number of misclassifications of the central quantile $q_0$ in favour of the upper one $q_{+1}$. Analysing results in Tables \ref{tab:weighted_mcc_horizon_10} and \ref{tab:weighted_mcc_horizon_50}, it is possible to note that there is no reason to place the current experiment in the same cluster as the Shallow-LSTM, as they are never statistically equivalent. It is clear that for different horizons $H_{\Delta \tau} |\Delta \tau \in \{10, 50, 100\}$, the model shows completely different behaviours. For horizon $H_{\Delta \tau} |\Delta \tau = 10$ it is statistically equivalent to the state-of-the-art methods which will be described in the next paragraph, but for $H_{\Delta \tau} |\Delta \tau \in \{50, 100\}$ there is no similarity to these models. This is the reason why the Attention-LSTM model is placed at the intersection of all clusters. \\

General performances for CNN-LSTM and MLP models are comparable to the ones for horizon $H_{\Delta \tau} |\Delta \tau = 10$ in the Shallow-LSTM model. The difference with the Shallow-LSTM experiment, making CNN-LSTM and MLP models state-of-the-art, must be searched in their ability to maintain stable, high performances throughout horizons $H_{\Delta \tau}$. Here too the higher values of weighted precision and recall for both the considered models, indicate the ability to correctly classify a significant number of samples associated with different target classes. Such an ability not only concerns higher level performance metrics, but is reflected in fine-grained per-class performance metrics as well. For different horizons $H_{\Delta \tau} |\Delta \tau \in \{10, 50, 100\}$, homogeneous class-specific weighted precision and recall are observed. The greater ability of these two models to correctly predict test samples is also shown by their highest values for Matthews Correlation Coefficient and Cohen's Kappa, indicating a better overlap between predictions and the ground truth. A statistical equivalence between these models throughout metrics and horizons arises from the results presented in Tables \ref{tab:weighted_mcc_horizon_10} and \ref{tab:weighted_mcc_horizon_50}.

\begin{table}[H]
\centering
\caption{Ranking representation of results from the Bayesian correlated t-test \cite{benavoli2017time} based on the MCC performance metric. Models on the same line indicate statistical equivalence and models in lower rows perform worse (statistically significant) than the ones in the upper rows.}
\label{tab:weighted_mcc_horizon_10}
\resizebox{12cm}{!}{%
\begin{tabular}{@{}ccc@{}}
\toprule
$\mathbf{H_{\Delta \tau} | \Delta \tau = 10}$ & $\mathbf{H_{\Delta \tau} | \Delta \tau = 50}$ & $\mathbf{H_{\Delta \tau} | \Delta \tau = 100}$ \\ \midrule
\rowcolor[HTML]{FFFFFF} 
\begin{tabular}[c]{@{}c@{}}Multilayer Perceptron \\ CNN - LSTM\end{tabular} & \begin{tabular}[c]{@{}c@{}}Multilayer Perceptron \\ CNN - LSTM\end{tabular} & \begin{tabular}[c]{@{}c@{}}Multilayer Perceptron \\ CNN - LSTM\end{tabular} \\ \midrule
\rowcolor[HTML]{FFFFFF} 
Self-Attention LSTM & \begin{tabular}[c]{@{}c@{}}Shallow LSTM\\ Multinomial Logistic Regression\end{tabular} & Multinomial Logistic Regression \\ \midrule
\rowcolor[HTML]{FFFFFF} 
\begin{tabular}[c]{@{}c@{}}Shallow LSTM\\ Multinomial Logistic Regression\end{tabular} & Self-Attention LSTM & Self-Attention LSTM \\ \midrule
\rowcolor[HTML]{FFFFFF} 
\begin{tabular}[c]{@{}c@{}} Naive Model \\ Random Model \end{tabular} & \begin{tabular}[c]{@{}c@{}} Naive Model \\ Random Model \end{tabular} & Shallow LSTM \\ \midrule
\rowcolor[HTML]{FFFFFF} 
 & & Random Model \\ \bottomrule
\end{tabular}
}
\end{table}

\begin{table}[H]
\centering
\caption{Ranking representation of results from the Bayesian correlated t-test \cite{benavoli2017time} based on the F-measure performance metric. Models on the same line indicate statistical equivalence and models in lower rows perform worse (statistically significant) than the ones in the upper rows.}
\label{tab:weighted_mcc_horizon_50}
\resizebox{12cm}{!}{%
\begin{tabular}{@{}ccc@{}}
\toprule
$\mathbf{H_{\Delta \tau} | \Delta \tau = 10}$ & $\mathbf{H_{\Delta \tau} | \Delta \tau = 50}$ & $\mathbf{H_{\Delta \tau} | \Delta \tau = 100}$ \\ \midrule
\rowcolor[HTML]{FFFFFF} 
\begin{tabular}[c]{@{}c@{}}Multilayer Perceptron \\ CNN - LSTM\\ Self-Attention LSTM\end{tabular} & \begin{tabular}[c]{@{}c@{}}Multilayer Perceptron \\ CNN - LSTM\end{tabular} & Multilayer Perceptron \\ \midrule
\rowcolor[HTML]{FFFFFF} 
\begin{tabular}[c]{@{}c@{}}Shallow LSTM\\ Multinomial Logistic Regression\end{tabular} & Shallow LSTM & \begin{tabular}[c]{@{}c@{}}CNN - LSTM\\ Multinomial Logistic Regression\end{tabular} \\ \midrule
\rowcolor[HTML]{FFFFFF} 
Naive Model & Multinomial Logistic Regression & Shallow LSTM \\ \midrule
\rowcolor[HTML]{FFFFFF} 
 Random Model & Self-Attention LSTM & Self-Attention LSTM \\ \midrule
\rowcolor[HTML]{FFFFFF} 
 & Naive Model & Naive Model \\ \midrule
\rowcolor[HTML]{FFFFFF} 
 & Random Model & Random Model \\ \bottomrule
\end{tabular}
}
\end{table}

\section{Conclusions} \label{conclusions}

In the present work, different Deep Learning models are applied to the task of price return forecasting in financial markets based on the Limit Order Book. LOBSTER data is used to train and test the Deep Learning models which are then analysed in terms of results, similarities between the models, performance ranking and dynamics-based model embedding. Hypotheses regarding the nature of the Limit Order Book and its dynamics are then discussed on the basis of model performances and similarities.

The three main contributions of the present work are summarised hereafter and directions for future work are suggested.

The Multinomial Logistic Regression model is solid throughout horizons $H_{\Delta \tau}$ but lacks the ability to produce complex features which could improve its performance as well as explicit dynamics modeling. Not all complex architectures are though able to outperform the Multinomial Logistic Regression model, such as the shallow LSTM model. This architecture incorporates the temporal dimension alone and does not significantly outperform the Logistic Regression, yielding a decrease in predictive power for longer horizons (i.e. $H_{\Delta \tau} |\Delta \tau \in \{50, 100\}$). The time dimension, upon which recurrent models are based, is also exploited by the Self-Attention LSTM model, which is augmented by the Self-Attention module. This allows to consider the whole context, while calculating the relevance of specific components. This shrewdness guarantees state-of-the-art performances (in line with the CNN-LSTM and MLP) for short-range horizons (i.e. $H_{\Delta \tau} |\Delta \tau = 10$).

This result then leads to the second consideration. It is clear that multiple levels of complexity in terms of return prediction exist in an LOB. There are at least two levels of complexity. The first one relates to short-range predictions (i.e. horizon $H_{\Delta \tau} |\Delta \tau = 10$). It is time dependent and can be well predicted by statically considering spatial dynamics which can be found in the immediate history (the context) related to the LOB state at tick time \textit{t}. The second level of complexity is related to longer-range predictions (i.e. horizons $H_{\Delta \tau} |\Delta \tau \in \{50, 100\}$) and multiple dimensions (temporal and spatial) must be taken into account to interpret it. The CNN-LSTM model, which explicitly models both dynamics, seems to penetrate deeper LOB levels of complexity guaranteeing stable and more accurate predictions for longer horizons too. This finding, combined with the results previously discussed, would lead to assert that space and time could be building blocks of the LOB's inner structure. This hypothesis is though partially denied by the statistically equivalent performance of the Multilayer Perceptron.

The last consideration hence follows from this. It is observed that a simple Multilayer Perceptron, which does not explicitly model temporal or spatial dynamics, yields statistically equivalent results to the CNN-LSTM model, the current state-of-the-art. According to these results, it is possible to conclude that both time and space are a good approximations of the underlying LOB's dimensions for the different prediction horizons $H_{\Delta \tau}$, but they should not be considered the real, necessary underlying dimensions ruling this entity and hence the market.

It is important to carry over some important considerations made throughout the paper which put the work into context and lead to future work. In the present work we investigate price change-based horizons, which hence correct for market volatlity. Future works should investigate the more difficult and noisy - but very insightful in practice -  task of predicting with ``real time'' horizons. The MLP used here is passed the same historical data as temporal models - per its structure it does not encode sequentiality (which the model could through infer), but we do not make any statements about the near LOB history being negligible. The considered stock falls into the category of large tick stocks which are vastly investigated in the literature and are know to be more suitable for ML-based predictions than small tick ones. Future work should investigate a now overdue framework for ML-based prediction in small tick stocks with state of the art performance comparable to that achieved on large tick stocks.

The present work has demonstrated how Deep Learning can serve a theoretical and interpretative purpose in financial markets. Future works should further explore Deep Learning-based theoretical investigations of financial markets and trading dynamics. The upcoming paper by the authors of this work presents an extension of the concepts in this work to Deep Reinforcement Learning and calls for further theoretical agent-based work in the field of high frequency trading and market making.

\section{Acknowledgements}
The authors acknowledge Riccardo Marcaccioli for the useful discussions and suggestions which have improved the quality of this work and sparked directions for future work. AB acknowledges support from the European Erasmus+ Traineeship Program. TA and JT acknowledge support from the EC Horizon 2020 FIN-Tech (H2020-ICT-2018-2 825215) and EPSRC (EP/L015129/1). TA also acknowledges support from ESRC (ES/K002309/1); EPSRC (EP/P031730/1).

%\bibliographystyle{abbrv} 
%\bibliography{references}
\printbibliography

\begin{appendices}

\section{Bayesian-based model clustering}\label{appendix: appendixA}
\begin{figure}[H]
    \centering
    \includegraphics[scale=0.25]{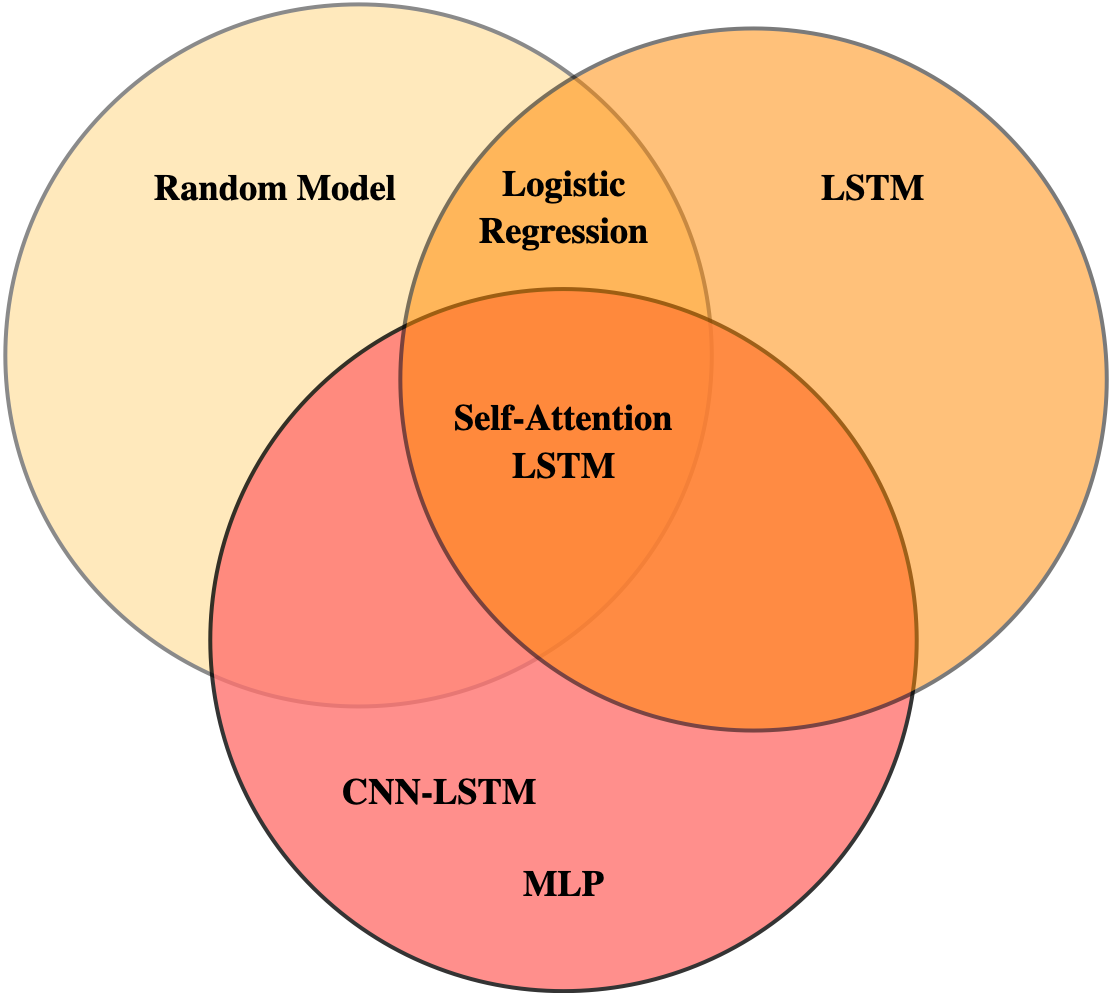}
    \caption{Schematic clustering solution for models presented in Section \ref{exps}. Similarities between models’ performances are tested using the Bayesian Correlated t-test described by Benavoli \textit{et al.} \cite{benavoli2017time}. Models contained in the same cluster component perform statistically equivalent based on MCC and weighted F-measure metrics. The higher intensity in cluster shading indicates an increase in models performances.}
    \label{fig:performances_clusters}
\end{figure}

\end{appendices}

\end{document}